\RequirePackage{fix-cm}
\documentclass[twocolumn,epjc3]{svjour3}
\smartqed  
\RequirePackage{graphicx}
\journalname{Eur. Phys. J. C}
\begin{document}

\title{Higgs boson, renormalization group, and naturalness in
cosmology
}


\author{A. O. Barvinsky\thanksref{e1,addr1},
A. Yu. Kamenshchik\thanksref{e2,addr2,addr3}, 
C. Kiefer\thanksref{e3,addr4},
A.~A.~Starobinsky\thanksref{e4,addr3}
\and
C. F. Steinwachs\thanksref{e5,addr4,addr5}}

\thankstext{e1}{e-mail: barvin@td.lpi.ru}
\thankstext{e2}{e-mail: kamenshchik@bo.infn.it}
\thankstext{e3}{e-mail: kiefer@thp.uni-koeln.de}
\thankstext{e4}{e-mail: alstar@landau.ac.ru}
\thankstext{e5}{e-mail: cst@thp.uni-koeln.de}

\institute{Theory Department, Lebedev
Physics Institute,
Leninsky Prospect 53, Moscow 119991, Russia \label{addr1}
\and
Dipartimento di Fisica  ed  Astronomia and INFN,
via Irnerio 46, 40126 Bologna, Italy \label{addr2}
\and
L. D. Landau Institute for
Theoretical Physics, Russian Academy of Sciences,  Moscow 119334, Russia\label{addr3}
\and
Institut f\"ur Theoretische Physik,
Universit\"at zu K\"oln, Z\"ulpicher Strasse 77,
50937 K\"oln, Germany\label{addr4}
\and
New address from September 1, 2012: University of Nottingham, University Park,
Nottingham, NG7 2RD, UK\label{addr5}}

\date{Received: date / Accepted: date}

\maketitle

\begin{abstract}
We consider the renormalization group improvement in the theory of the
Standard Model (SM) Higgs boson playing the role of an inflaton with a
strong non-minimal coupling to gravity. At the one-loop level with the
running of constants taken into account, it leads to a range of the
Higgs mass that is entirely determined by the lower WMAP bound on the
cosmic microwave background (CMB) spectral index. We find that the SM
phenomenology is sensitive to current cosmological data, which
suggests to perform more precise
CMB measurements as a SM test complementary to the LHC program. By
using the concept of a field-dependent cutoff, we show the naturalness of
the gradient and curvature expansion in this model within the
conventional perturbation theory range of the SM. We also discuss the
relation of these results to two-loop calculations and the
limitations of the latter caused by parametrization and gauge
dependence problems.

\end{abstract}
\section{Introduction}

The announcement of the Higgs boson discovery at the LHC in a narrow mass
range close to 125 GeV \cite{CERN} draws attention to the Higgs
inflation model for the physics of the early Universe. An obvious rationale behind this is the anticipation that cosmological observations can comprise SM
tests complimentary to collider experiments.

While the relatively old work \cite{we-scale} had suggested that, due
to quantum effects, inflation depends not only on the inflaton--graviton
sector of the system, but is strongly effected by the Grand Unified
Theory (GUT)
contents of the particle model, the series of papers
\cite{BezShap,we,BezShap1,Wil,BezShap3} implemented this idea into the
context of the SM, with the Higgs field playing the role of an inflaton. This
has led to new interest in a once rather popular model
\cite{non-min,we-scale,BK,KomatsuFutamase} with the Lagrangian
of the graviton--inflaton sector given by
    \begin{eqnarray}
    &&L(g_{\mu\nu},\Phi)=
    \frac12\left(M_{\rm P}^2+\xi|\Phi|^2\right)R
    -\frac{1}{2}|\nabla\Phi|^{2}
    -V(|\Phi|),                     \label{inf-grav}\\
    &&V(|\Phi|)=
    \frac{\lambda}{4}(|\Phi|^2-v^2)^2,\,\,\,\,
    |\Phi|^2=\Phi^\dag\Phi,
    \end{eqnarray}
where $\Phi$ is a scalar field multiplet, whose expectation value plays
the role of an inflaton and which has a strong non-minimal curvature
coupling with $\xi\gg 1$. Here, $M_{\rm P}=m_{\rm P}/\sqrt{8\pi}\approx
2.4\times 10^{18}$ GeV is a
reduced Planck mass, $\lambda$ is a quartic self-coupling of
$\Phi$, and $v$ is a symmetry breaking scale.

The motivation for this model was  based on the observation
\cite{non-min} that the problem of an exceedingly small quartic
coupling $\lambda\sim 10^{-13}$, as dictated by the amplitude of
primordial CMB perturbations \cite{CMB}, can be circumvented by using
a non-minimally coupled inflaton with a large value of $\xi$. Later,
this model with the GUT-type sector of matter fields was used to
generate initial conditions for inflation \cite{we-scale} within the
concept of the no-boundary \cite{noboundary} and tunneling
cosmological state \cite{tunnel}. The quantum evolution with these
initial data was considered in \cite{BK,efeqmy}. There, it was shown
that quantum effects are critically important for this scenario.

A similar theory, but with the SM Higgs boson $\Phi$ playing the role
of an inflaton instead of the abstract GUT setup of \cite{we-scale,BK},
was suggested in \cite{BezShap}. In particular, it was advocated that
the corresponding CMB data are consistent with the WMAP observations in
the tree-level approximation of the theory.

The further history of
this non-minmally coupled Higgs inflation model was as follows.
The methods of \cite{we-scale,BK,efeqmy} were used to extend
the predictions in this model to the one-loop level \cite{we}. This
has led immediately to the lower bound on the Higgs mass $M_{\rm H}\approx 230$
GeV, originating from the observational restrictions on the CMB
spectral index \cite{we}. However, this conclusion did not take into
account $O(1)$ effects of the renormalization group (RG) running,
which qualitatively change the situation. This was nearly
simultaneously observed in \cite{Wil} and in \cite{BezShap1}, where the RG
improvement of the one-loop results of \cite{we} has decreased the
lower bound on the Higgs mass to about $135$ GeV.

Quantitatively, this result was confirmed in our paper \cite{RGHiggs},
where we suggested the RG improvement of our one-loop results in
\cite{we} and found a range of the Higgs mass that is compatible with the
CMB; both the lower and upper boundary of this range
are determined by the lower WMAP bound on the CMB
spectral index, $n_s\approx 0.94$. The predictions of this model have
also been extended to the two-loop approximation \cite{Wil,BezShap3},
which has led to a reduction of
the lower bound on the Higgs mass range by about 10
GeV, which then nearly coincides with the recently announced value
of about 125 GeV.

Simultaneously with the papers advocating Higgs inflation, including
in particular its supersymmetric extension \cite{supersym,KS12}, there
arose a number of objections to this model. Apart from the strong
assumption that no ``new physics" is present between electroweak (EW) and
inflation scales, it was criticized on the basis that the predictions for Higgs
inflation rely on perturbation
theory, which is only valid below the strong coupling scale.
The reason for this criticism is that, for flat
space perturbation theory with a vanishing Higgs field background, this
scale turns out to be $M_{\rm P}/\xi$, which is much lower than the
inflation scale $M_{\rm P}/\sqrt\xi$
\cite{BurgLeeTrott1,Barbon,BMSS}, rendering the application of
perturbation theory questionable. Moreover,
the multi-component nature of the Higgs
field leads to the impossibility of canonically normalizing all
its components in the Einstein frame \cite{BurgLeeTrott1,Hertzberg,Kaiser} -- the
parametrization heavily employed in \cite{Wil,BezShap3}.

The Higgs inflation model was also discussed  in \cite{Percacci} in the approach of asymptotically safe gravity \cite{Weinberg,asympsafegrav}.
There,
however, one does
not use the model (\ref{inf-grav}) with large $\xi$, but instead
exploits a rather miraculous numerological observation -- a certain
relation between the EW instability in the SM and
the Planck scale \cite{asympsafe,BezShap4}.\footnote{The fixed point of the
  running coupling $\lambda(t)$ occurs very close to the Planck
  scale $t_{\rm P}$ with $\lambda(t_{\rm P})=0$.}

In spite of all these objections, the remarkable conformity of the LHC
tests and the Higgs mass range compatible with the CMB data makes this
model extremely attractive. What remains disturbing is that
this agreement between various calculations
\cite{Wil,BezShap3,RGHiggs},   does
not exceed the $O(1)$ precision. In fact, it does not go beyond the
order-of-magnitude observation made in the first paper \cite{we},
which relates the WMAP
data for the red tilt of the CMB spectrum with the value of the Higgs
mass. If the Higgs boson indeed exists with a mass near 125 GeV, then,
according to \cite{Wil,BezShap3}, the two-loop approximation will be really
necessary to bring Higgs inflation in accordance with EW
physics. However, the
difference of 10~GeV between the RG improved one-loop result of
\cite{BezShap1,RGHiggs} and the two-loop results of
\cite{Wil,BezShap3} comprises a ten percent correction and thus, at least
naively, greatly exceeds the order-of-magnitude correction
$\lambda/16\pi^2\sim 0.01$ that one would expect in the next order of
the loop expansion. Despite this alerting fact, the authors of
\cite{BezShap3,BezShap4} went beyond a simple estimate of the Higgs
mass range, leading to 126.1 GeV $\leq M_{\rm H} \leq$ 193.9 GeV, and
meticulously
established various types of uncertainties contributing to its error
bars (uncertainties in the top quark mass, in the running of strong
coupling constant, theoretical uncertainties in multi-loop
contributions, etc.). Moreover, the original motivation of deriving
the Higgs mass from inflation
\cite{BezShap1,BezShap3} has shifted to an analysis of the
EW vacuum stability within the asymptotic safety hypothesis
\cite{BezShap4}, one of the main conclusions of this work being that
these stability arguments automatically provide a consistent Higgs
inflation scenario. According to \cite{BezShap4}, this conclusion
depends very
weakly on the way how the SM is embedded into the gravitational context
if the Higgs mass belongs to a certain range close to the one
suggested in \cite{BezShap3}.

It is hard to agree with this statement on the universality of such an
embedding, because the non-minimal coupling of the Higgs multiplet
drastically changes the RG running of the SM coupling constants. Moreover,
from the viewpoint of Higgs inflation, the meticulous bookkeeping of error
bars in the two-loop approximation of \cite{BezShap3} and the two-loop
approximation itself essentially exceeds the available precision of the
current CMB data and, what is much more important, the theoretical
status of  quantum corrections to physical observables in inflation
theory. If the cornerstone of the cosmological tests of the SM is to
measure correlations of cosmological observables, then these
observables should be defined in a generally covariant manner. The
problem of their gauge-invariant definition on the background of a {\em fiducial} FRW
metric has been resolved long ago in the linear order (see
\cite{perturbations1} and references therein) and beyond
\cite{perturbations2}, but in our case this fiducial metric becomes a
dynamical quantum mean field subject to effective equations of motion,
and this drastically changes the situation. Via this mean field, an
essential gauge and parametrization dependence enters the definition
of cosmological observables and this problem has not yet been resolved. In particular, the debate of the Jordan
frame versus the Einstein frame calculations
\cite{we,BezShap1,RGHiggs,BezShap3} is a part of this parametrization
dependence story. Even
the one-loop approximation might be sensitive to  nonlinear
corrections, whereas in the two-loop approximation these corrections
enter the game in full weight. It is important that relevant
ambiguities  not only affect the gravitational sector of the model, but also
arise in the purely SM sector leading to  dependence on the
Higgs-multiplet parametrization.\footnote{The effective
  potentials of the Higgs multiplet calculated within its Cartesian
  parametrization and spherical parametrization are
  different off-shell, because the latter does not  include the
  contribution of Nambu--Goldstone (angular) modes at all. This strongly
  affects solutions of effective equations of motion, in which a small
  but nonzero slope of the effective potential provides a slow-roll
  regime of  inflationary dynamics. This is one of the sources of
  discrepancies between the results of \cite{RGHiggs} and
  \cite{BezShap1,BezShap3}, which arise already in the one-loop
  approximation.}

For these reasons, calculations beyond the one-loop approximation do
not seem useful, unless the problems of the above type get a consistent
resolution. Therefore, in our discussion of the Higgs inflation below
we restrict ourselves to the one-loop results and their RG
improvement. In Sect. 2, we begin with the review of the non-minimal
Higgs inflation model and its CMB parameters when the tree-level
potential of the Higgs-inflaton field is replaced by the one-loop
effective potential. In Sects. 3 and 4, we recapitulate the results of
the RG improvement in this model with the running coupling constants
interpolating between the EW and inflation scales, which leads
to a definite range of the Higgs mass compatible with the current CMB
data \cite{RGHiggs}.
In Sect. 5, we analyse the validity of the gradient and curvature
expansion and demonstrate its naturalness with the {\em background
  dependent} cutoff during the whole inflationary and
post-inflationary evolution. The naturalness of the loop expansion for the
effective potential was shown in \cite{BMSS} in the Einstein frame of
the theory, which leaves us the hope that the potential
parametrization (or frame) independent formulation of the quantum
Higgs inflation model will guarantee the naturalness of this expansion,
too. In Sect. 6, we summarize and discuss our conclusions and sketch
 future prospects of this model within the ongoing  quest for unifying the Higgs
boson physics with   studies of the large-scale structure of the Universe.

\section{CMB parameters in the non-minimal
        Higgs inflation model}

The usual understanding of non-renormalizable theories is that
renormalization of higher-dimensional operators does not affect the
renormalizable sector of low-dimensional operators, because the former
ones are suppressed by powers of a cutoff -- the Planck mass $M_{\rm
  P}$ \cite{Weinberg}. Therefore, beta functions of the Standard Model
sector are not expected to be modified by gravitons.

The situation
with the non-minimal coupling is more subtle. Due to the mixing of the
Higgs scalar field with the longitudinal part of gravity in the
kinetic term of the Lagrangian (\ref{inf-grav}), an obvious
suppression of pure graviton loops by the effective Planck mass,
$M_{\rm P}^2+\xi\varphi^2\gg M_{\rm P}^2$, proliferates for large
$\xi$  to the sector of the Higgs field, so that certain parts of the
beta functions are strongly damped by a large $\xi$
\cite{our-ren,Wil}.
Therefore, a special combination of coupling constants
$\mbox{\boldmath$A$}$ which we call {\em anomalous scaling}
\cite{we-scale} becomes very small  that decreases the Higgs mass 
lower bound and makes it compatible with the CMB observational data.  
The importance of this quantity
follows from the fact observed in \cite{we-scale,BK,we} that, due
to large $\xi$, quantum effects and their CMB manifestation are
universally determined by $\mbox{\boldmath$A$}$. The nature of
this quantity is as follows.

Let the model contain in addition to (\ref{inf-grav}) also a set of
scalar fields $\chi$, vector gauge bosons $A_\mu$ and spinors $\psi$,
which have an interaction with $\Phi$ dictated by the local gauge
invariance. If we denote by $\varphi$ the inflaton -- the only nonzero
component of the mean value of $\Phi$ in the cosmological state, then
the quantum effective action of the system takes the generic form
\begin{eqnarray}
    &&S[g_{\mu\nu},\varphi]=\int d^{4}x\,g^{1/2}
    \left(-V(\varphi)+U(\varphi)\,R(g_{\mu\nu})\right.\nonumber \\-
    &&\left.\frac12\,G(\varphi)\,(\nabla\varphi)^2+...\right),
       \label{effaction}
    \end{eqnarray}
where $V(\varphi)$, $U(\varphi)$ and $G(\varphi)$ are the
coefficients of the derivative expansion, and we disregard the
contribution of higher-derivative operators that are negligible in the
slow-roll approximation of the inflation theory. In this
approximation, the dominant quantum contribution to these
coefficients comes from the heavy massive sector of the model. In
particular, the masses of the physical particles and Goldstone modes
$m(\varphi)$, generated by their quartic, gauge and Yukawa couplings
with $\varphi$, give rise to the Coleman--Weinberg potential -- the
one-loop contribution to the effective potential $V$ in
(\ref{effaction}). Since $m(\varphi)\sim\varphi$, for large
$\varphi$ this potential is given by the following sum of boson and
fermion contributions:
   \begin{eqnarray}
    &&V^{\rm 1-loop}(\varphi)=\sum_{
    \rm particles}
    (\pm 1)\,\frac{m^4(\varphi)}{64\pi^2}
    \,\ln\frac{m^2(\varphi)}{\mu^2}\nonumber \\
    &&=\frac{\lambda\mbox{\boldmath$A$}}{128\pi^2}
    \,\varphi^4
    \ln\frac{\varphi^2}{\mu^2}+...  \label{Aviamasses}
    \end{eqnarray}
and thus determines the dimensionless coefficient
$\mbox{\boldmath$A$}$ -- the anomalous scaling associated with the
normalization scale $\mu$ in (\ref{Aviamasses}). Moreover, for $\xi\gg
1$ it is mainly this quantity and the dominant quantum correction to
$U(\varphi)$ \cite{RGHiggs},
   \begin{eqnarray}
   U^{\rm 1-loop}(\varphi)=
    \frac{3\xi\lambda}{32\pi^2}\,\varphi^2
    \ln\frac{\varphi^2}{\mu^2}+...\, ,     \label{U1loop}
    \end{eqnarray}
which determine the quantum rolling force in the effective equation of the
inflationary dynamics \cite{BK,efeqmy} and  which yield the parameters of
the CMB generated during inflation \cite{we}.

Inflation and the CMB are easy to analyse in the Einstein frame of
fields, denoted by $\hat g_{\mu\nu}$, $\hat\varphi$, in which the action $\hat
S[\hat g_{\mu\nu},\hat\varphi]=S[g_{\mu\nu},\varphi]$ has a minimal
coupling $\hat U=M_{\rm P}^2/2$, a canonically normalized inflaton
field $\hat G=1$, and the new inflaton potential $\hat{V}=M_{\rm P}^4
V(\varphi)/4U^2(\varphi)$.\footnote{The Einstein and Jordan frames
are related by the equations $\hat
g_{\mu\nu}=2U(\varphi)g_{\mu\nu}/M_{\rm P}^2$, $
(d\hat\varphi/d\varphi)^2=M_{\rm P}^2(GU+3U'^2)/2U^2$.} At the
inflationary scale with $\varphi>M_{\rm P}/\sqrt{\xi}\gg v$ and $\xi\gg
1$, this potential reads
        \begin{eqnarray}
        \hat{V}=\frac{\lambda
        M_{\rm P}^4}{4\,\xi^2}\,\left(1-\frac{2M_{\rm P}^2}{\xi\varphi^2}+
        \frac{\mbox{\boldmath$A_I$}}{16\pi^2}
        \ln\frac{\varphi}{\mu}\right),            \label{hatVbigphi}
        \end{eqnarray}
where the parameter $\mbox{\boldmath$A_I$}$ represents the anomalous
scaling (see (\ref{A0}) below) modified by the quantum correction to the
non-minimal curvature coupling (\ref{U1loop}),
       \begin{eqnarray}
        \mbox{\boldmath$A_I$}&=&\mbox{\boldmath$A$}-12\lambda\nonumber\\
        &=&
        \frac3{8\lambda}\Big(2g^4 +
        \big(g^2 + g'^2\big)^2- 16y_t^4 \Big)-6\lambda.  \label{AI}
        \end{eqnarray}
This quantity -- which we shall call {\em inflationary anomalous
scaling} -- enters the expressions for the slow-roll parameters,
       \begin{eqnarray}
       \hat\varepsilon \equiv
       \frac{M_{\rm P}^2}2\left(\frac1{\hat V}\frac{d\hat
           V}{d\hat\varphi}\right)^2,\quad
       \hat\eta\equiv \frac{M_{\rm P}^2}{\hat V}
       \frac{d^2\hat V}{d\hat\varphi^2},
       \end{eqnarray}
and ultimately determines all the inflation characteristics. In
particular, the smallness of $\hat\varepsilon$ yields the range of the
inflationary stage $\varphi>\varphi_{\rm end}$, terminating at a value
of $\hat\varepsilon$ which we choose to be
$\hat\varepsilon_{\rm end}=3/4$.
Under the natural assumption that perturbation expansion is
applicable for $\mbox{\boldmath$A_I$}/64\pi^2\ll 1$,
the inflaton value at the exit
from inflation then equals $\varphi_{\rm end}\simeq 2M_{\rm P}/\sqrt{3\xi}$.
The value of
$\varphi$ at the beginning of the inflation stage of duration $N$ in
units of the e-folding number then reads \cite{we}
    \begin{eqnarray}
    &&\varphi^2=\frac{4N}3\frac{M_{\rm
        P}^2}{\xi}\frac{e^x-1}x, \label{xversusvarphi}\\
    &&x\equiv\frac{N
    \mbox{\boldmath$A_I$}}{48\pi^2},           \label{x}
    \end{eqnarray}
where the special parameter $x$ directly involves the anomalous
scaling $\mbox{\boldmath$A_I$}$.

This relation determines the Fourier power spectrum for the scalar
metric perturbation $\zeta$,
$\Delta_{\zeta}^2(k) \equiv \langle k^3\zeta_{{\bf k}}^2\rangle
= \hat V/24\pi^2M_{\rm P}^4\hat\varepsilon$,
where the right-hand side is taken
at the  first horizon crossing, $k=aH$, relating the comoving
perturbation wavelength $k^{-1}$ to the e-folding number $N$,
    \begin{eqnarray}
    \Delta_{\zeta}^2(k)=
    \frac{N^2}{72\pi^2}\,\frac\lambda{\xi^2}\,
    \left(\frac{e^x-1}{x\,e^x}\right)^2.       \label{zeta}
    \end{eqnarray}
The CMB spectral index $n_s\equiv 1+d\ln\Delta_{\zeta}^2/d\ln
k=1-6\hat\varepsilon+2\hat\eta$ and the tensor to scalar ratio
$r=16\hat\varepsilon$ correspondingly read as\footnote{Note that for
$|x|\ll 1$ these predictions exactly coincide with those \cite{S83}
of the $f(R)=M_{\rm P}^2(R+R^2/6M^2)/2$ inflationary model \cite{S80} with
the scalar particle (scalaron) mass $M=M_{\rm P}\sqrt \lambda/\sqrt 3
\xi$.}
    \begin{eqnarray}
    &&n_s=
    1-\frac{2}{N}\, \frac{x}{e^x-1}~,           \label{ns}\\
    &&r=\frac{12}{N^2}\,
    \left(\frac{x e^x}{e^x-1}\right)^2~.          \label{r}
    \end{eqnarray}
Therefore, with the spectral index constraint $0.948 <n_s(k_0)<0.986$
(the combined WMAP+SPT+BAO+$H_0$ data at the $2\sigma$ confidence level
with the pivot point $k_0=0.002$
Mpc$^{-1}$ \cite{WMAP,SPT}  corresponding to $N\simeq 60$) , these
relations immediately give the range
$-12<\mbox{\boldmath$A_I$}<14$
for the inflationary anomalous scaling \cite{we}.

In the SM, $\mbox{\boldmath$A$}$ is
expressed in terms of the masses of the heaviest particles -- $W^\pm$
boson, $Z$ boson and top quark,
    \begin{eqnarray}
    &&m_W^2=\frac14\,g^2\,\varphi^2,\;\;
    m_Z^2=\frac14\,(g^2+g'^2)\,\varphi^2,\nonumber\\
    &&m_t^2=\frac12\,y_t^2\,
    \varphi^2,                       \label{masses}
    \end{eqnarray}
and the mass of the three Goldstone modes
$m_G^2=V'(\varphi)/\varphi=\lambda(\varphi^2-v^2)\simeq
\lambda\varphi^2$. Here, $g$ and $g'$ are the $SU(2)\times U(1)$
gauge couplings, $g_s$ is the $SU(3)$ strong coupling, and $y_t$ is
the Yukawa coupling for the top quark. At the inflation stage, the
Goldstone mass-squared $m_G^2$ is non-vanishing, in contrast to its zero
on-shell value in the EW vacuum \cite{WeinbergQFT}. Therefore, Eq.
(\ref{Aviamasses}) gives the expression
   \begin{equation}
    {\mbox{\boldmath $A$}} =
    \frac3{8\lambda}\Big(2g^4 +
    \big(g^2 + g'^2\big)^2- 16y_t^4 \Big)+6\lambda.   \label{A0}
    \end{equation}
In the conventional range of the Higgs mass 115 GeV$\leq M_{\rm H}\leq$
180 GeV \cite{particle}, this quantity lies at the EW scale
in  the range $-48<\mbox{\boldmath$A$}<-20$, which strongly
contradicts the CMB range given above.

However, the RG running of coupling constants is strong enough and
drives ${\mbox{\boldmath $A$}}$  to the CMB compatible range at the
inflation scale. Below we show that the formalism of \cite{we} stays
applicable but with the EW ${\mbox{\boldmath $A$}}$ replaced
by the running ${\mbox{\boldmath $A$}}(t)$, where $t=\ln(\varphi/\mu)$
is the running scale of the renormalization group (RG) improvement of
the effective potential \cite{ColemanWeinberg}.

\section{RG improvement}

According to the Coleman--Weinberg technique \cite{ColemanWeinberg},
the one-loop RG improved effective action has the form
(\ref{effaction}), with
    \begin{eqnarray}
    &&V(\varphi)=
    \frac{\lambda(t)}{4}\,Z^4(t)\,\varphi^4,  \label{RGeffpot}\\
    &&U(\varphi)=
    \frac12\Big(M_{\rm P}^2
    +\xi(t)\,Z^2(t)\,\varphi^{2}\Big),      \label{RGeffPlanck}\\
    &&G(\varphi)=Z^2(t).            \label{phirenorm1}
    \end{eqnarray}
Here, the running scale $t=\ln(\varphi/M_t)$ is normalized at the top quark mass $\mu=M_t$ (we denote physical (pole) masses by capital letters in contrast to running masses, see (\ref{masses}) above).\footnote{Application of the Coleman--Weinberg technique removes the ambiguity in the choice of the RG scale in cosmology -- an issue discussed in \cite{Woodard}.} The running couplings $\lambda(t)$,
$\xi(t)$ and the field renormalization $Z(t)$ incorporate a summation of
powers of logarithms and belong to the solution of the RG equations
    \begin{eqnarray}
    &&\frac{d g_i}{d t}
    =\beta_{g_i},\,\,\,\,\frac{dZ}{d t}
    =\gamma Z                   \label{renorm0}
    \end{eqnarray}
for the full set of coupling constants
\begin{eqnarray*}
& & g_i=(\lambda,\xi,g,g',g_s,y_t)
\end{eqnarray*}
in the ``heavy'' sector of the model
with the corresponding beta functions $\beta_{g_i}$ and the anomalous
dimension $\gamma$ of the Higgs field.

An important subtlety for these $\beta$ functions is the effect of the
non-minimal curvature coupling of the Higgs field. For large $\xi$, the
kinetic term of the tree-level action has a strong mixing between the
graviton $h_{\mu\nu}$ and the quantum part of the Higgs field $\sigma$
on the background $\varphi$. Symbolically, it has the structure
\[ (M_{\rm P}^2+\xi^2\varphi^2)h\nabla\nabla
h+\xi\varphi\sigma\nabla\nabla h+\sigma\triangle\sigma,\] which yields
a propagator whose elements are suppressed by a small $1/\xi$-factor
in all blocks of the $2\times2$ graviton-Higgs sector. For large
$\varphi\gg M_{\rm P}/\sqrt\xi$, the suppression of pure graviton
loops is, of course, obvious because  the effective Planck mass
squared strongly exceeds the Einstein one, $M_{\rm
  P}^2+\xi\varphi^2\gg M_{\rm P}^2$. Due to the mixing, this
suppression proliferates to the full graviton-Higgs sector of the
theory and yields the Higgs propagator
$s(\varphi)/(\triangle-m_H^2)$, which contains the suppression factor
$s(\varphi)$ given by
    \begin{eqnarray}
    s(\varphi)=
    \frac{M_{\rm P}^2+\xi\varphi^2}
    {M_{\rm P}^2+(6\xi+1)\xi\varphi^2}.         \label{s}
    \end{eqnarray}

This mechanism \cite{our-ren,BK,efeqmy} modifies the beta functions of
the SM sector \cite{Wil} at high energy scales because the factor
$s(\varphi)$, which is close to one at the EW scale $v\ll M_{\rm
  P}/\xi$, is very small for $\varphi\gg M_{\rm P}/\sqrt\xi$, $s\simeq
1/6\xi$. Such a modification justifies, in fact, the extension beyond
the scale $M_{\rm P}/\xi$ which is interpreted in \cite{BurgLeeTrott,Barbon} as
a natural validity cutoff of the theory.\footnote{\label{footnote4}The cutoff $M_{\rm P}/\xi\ll M_{\rm P}$ of \cite{BurgLeeTrott,Barbon} applies to energies (momenta) of scattering processes in flat spacetime with a small EW value of $\varphi$. For the inflation stage on the background of a large and alsmost constant $\varphi$, this cutoff gets modified due to the increase in the effective Planck mass
$M_{\rm P}^2+\xi\varphi^2\gg M_{\rm P}^2$ (and the associated decrease of the $s$-factor (\ref{s}) -- resummation of terms treated otherwise as perturbations in \cite{BurgLeeTrott}). As was claimed in \cite{BMSS}, such a background dependent cutoff scale remains higher than the dynamical scale
throughout the whole history of the Universe, including inflation, reheating and low-energy physics in the present Universe 
apart from a short period during the middle stage of  reheating when $\varphi$ is oscillating and is of the order of $M_P/\xi$. But in the renormalization group approach we 
consider the sequence of background configurations with an almost constant scalar field $\varphi$ and need not follow the actual evolution of $\varphi(t)$ through the history of the Universe (see the discussion in Sects. 5 and 6 below).}

There is another important subtlety with the modification of beta
functions, which was disregarded in \cite{Wil} (and in the first
version of \cite{RGHiggs}). Goldstone modes, in contrast to the
Higgs particle, do not have  mixing with gravitons in the kinetic term
\cite{BezShap3}. Therefore, their contribution is not suppressed
by the $s$-factor of the above type. Separation of Goldstone
contributions from the Higgs contributions leads to the following
modification of the one-loop beta functions, which strongly
differs from that of \cite{Wil} (cf. also \cite{Clarcketal}):
    \begin{eqnarray}
    &&\beta_{\lambda} = \frac{\lambda}{16\pi^2}
    \left(18s^2\lambda
    +{\mbox{\boldmath $A$}}(t)\right)
    -4\gamma\lambda,                           \label{beta-lambda}\\
    &&\beta_{\xi} =
    \frac{6\xi}{16\pi^2}(1+s^2)\lambda
    -2\gamma\xi,                 \label{beta-xi}\\
    &&\beta_{y_t} = \frac{y_t}{16\pi^2}
    \left(-\frac{2}{3}g'^2
    - 8g_s^2 +\left(1+\frac{s}2\right)y_t^2\right)
    -\gamma y_t,                                    \label{beta-y}\\
    &&\beta_{g} = -\frac{39 - s}{12}
    \frac{g^3}{16\pi^2},                     \label{beta-g}\\
    &&\beta_{g'} =
    \frac{81 + s}{12} \frac{g'^3}{16\pi^2},  \label{beta-g1}\\
    &&\beta_{g_s} =
    -\frac{7 g_s^3}{16\pi^2}.                    \label{beta-gs}
    \end{eqnarray}
Here, the anomalous dimension $\gamma$ of the Higgs field  is given by
the standard expression in the Landau gauge,
    \begin{eqnarray}
    \gamma=\frac1{16\pi^2}\left(\,\frac{9g^2}4
    +\frac{3g'^2}4 -3y_t^2\right),                  \label{gamma}
    \end{eqnarray}
the anomalous scaling ${\mbox{\boldmath $A$}}(t)$ is defined by
(\ref{A0}),  and we have retained only the leading terms in $\xi\gg 1$. 
In what follows,
it
will be important  that this anomalous scaling
contains the Goldstone contribution $6\lambda$, so that the full
$\beta_\lambda$ in (\ref{beta-lambda}) has a $\lambda^2$-term
unsuppressed by $s(\varphi)$ at large scale $t=\ln(\varphi/\mu)$.

The inflationary stage in units of Higgs field e-fold\-ings is very
short, which allows one to use an approximation linear in $\Delta
t\equiv t-t_{\rm end}= \ln(\varphi/\varphi_{\rm end})$, where the
initial data point is chosen at the end of inflation $t_{\rm end}$.
Therefore, for the beta functions (\ref{beta-lambda}) and
(\ref{beta-xi}) with $s\ll 1$ we have
    \begin{eqnarray}
    &&\lambda(t) = \lambda_{\rm end}\left(1
    - 4\gamma_{\rm end}\Delta t
    +\frac{\mbox{\boldmath $A$}(t_{\rm end})}{16\pi^2}\,
    \Delta t\right),                             \label{lambda-lin}\\
    &&\xi(t) = \xi_{\rm end}\Big(1
    -2\gamma_{\rm end}\Delta t
    +\frac{6\lambda_{\rm end}}{16\pi^2}\Delta t\Big),     \label{xi-lin}
    \end{eqnarray}
where $\lambda_{\rm end}$, $\gamma_{\rm end}$, $\xi_{\rm end}$ are
determined at $t_{\rm end}$, and ${\mbox{\boldmath $A$}}_{\rm
end}={\mbox{\boldmath $A$}}(t_{\rm end})$ is the particular value
of the running anomalous scaling (\ref{A0}) at the end of inflation.

On the other hand, the RG improvement of the effective action
(\ref{RGeffpot})--(\ref{phirenorm1}) implies that this action
coincides with the tree-level action  for a new field
$\tilde{\varphi}=Z(t)\varphi$ with running couplings as functions of
$t=\ln(\varphi/\mu)$ (the running of $Z(t)$ is slow and affects only
the multi-loop RG improvement). Then, in view of
(\ref{RGeffpot})--(\ref{RGeffPlanck}), at the inflationary stage the RG improved potential takes
 the form of the one-loop potential
(\ref{hatVbigphi}) for the field $\varphi$ with a particular choice of
the normalization point $\mu=\varphi_{\rm end}$ and all the couplings
replaced by their running values taken at $t_{\rm end}$. Therefore,
the formalism of \cite{we} can be directly applied to find the 
parameters of the model relevant for the CMB data, 
which now turn out to be determined by the
running anomalous scaling ${\mbox{\boldmath $A_I$}}(t)$ taken at
$t_{\rm end}$ .

In contrast to the inflationary stage, the post - inflationary running
is very large and requires numerical simulation \cite{RGHiggs}. 
One of the reasons for this is that the assumption of an almost 
constant $\varphi$ background is broken at this stage. 
We fix
the $t=0$ initial conditions for the RG equations (\ref{renorm0}) at
the top quark scale $M_t =171$ GeV. For the constants $g,g'$ and
$g_s$, they read \cite{particle}
    \begin{equation}
    g^2(0) = 0.4202,\  g'^2(0) = 0.1291,
    \ g_s^2(0) = 1.3460,                          \label{initial}
    \end{equation}
where $g^2(0)$ and $g'^2(0)$ are obtained by a simple one-loop RG flow
from the conventional values of $\alpha(M_Z)\equiv
g^2/4\pi=0.0338$, $\alpha'(M_Z)\equiv g'^2/4\pi=0.0102$ at
$M_Z$-scale, and the value $g_s^2(0)$ at $M_t$ is generated by the
numerical program of \cite{QCDfromZtotop}.
The analytical algorithm of transition between different scales for $g_s^2$
was presented in  \cite{DV}.
For the Higgs
self-interaction constant $\lambda$ and for the Yukawa top quark
interaction constant $y_t$, the initial conditions are determined by
the pole mass matching scheme originally developed in \cite{top} and
presented in the Appendix of \cite{espinosa}.

The initial condition $\xi(0)$ follows from the CMB normalization
(\ref{zeta}), $\Delta_{\zeta}^2\simeq 2.5\times 10^{-9}$, at the
pivot point $k_0=0.002$ Mpc$^{-1}$ \cite{WMAP}, which we choose to
correspond to $N\simeq 60$
(this value is close to that proposed in \cite{BezGorb}).
This yields the following estimate for
the ratio of coupling constants,
    \begin{equation}
\frac{1}{Z_{\rm in}^2}\frac{\lambda_{\rm in}}{\xi^2_{\rm in}}
    \simeq 0.5\times 10^{-9}
    \left(\frac{x_{\rm in}\,\exp x_{\rm in}}
    {\exp x_{\rm in}-1}\right)^2 ,             \label{final}
    \end{equation}
at the moment of the first horizon crossing for $N=60$, which we call
the ``beginning'' of inflation and label by $t_{\rm
in}=\ln(\varphi_{\rm in}/M_t)$ with $\varphi_{\rm in}$ defined by
(\ref{xversusvarphi}). Thus, the RG equations (\ref{renorm0}) for the
six couplings $(g,g',g_s,y_t,\lambda,\xi)$ with five initial
conditions and the final condition for $\xi$ uniquely determine the
needed RG flow.

The RG flow covers also the inflationary stage from the
chronological end of inflation $t_{\rm end}$  to $t_{\rm in}$. At the
end of inflation we choose the value of the slow roll
parameter $\hat\varepsilon=3/4$, and $\varphi_{\rm end}\equiv M_t
e^{t_{\rm end}}\simeq M_{\rm P}\sqrt{4/3\xi_{\rm end}}$. Thus, the
duration of inflation in units of inflaton field e-foldings
$t_{\rm in}-t_{\rm end}=\ln(\varphi_{\rm in}/\varphi_{\rm
end})\simeq\ln N/2\sim 2$ \cite{RGHiggs} is very short relative to the
post-inflationary evolution $t_{\rm end}\sim 35$. The
approximation linear in the logarithms implies the bound $|{\mbox{\boldmath
$A_I$}}(t_{\rm end})|\Delta t/16\pi^2\ll 1$, which in view of
$\Delta t<t_{\rm in}-t_{\rm end}\simeq \ln N/2$ holds for
$|{\mbox{\boldmath $A_I$}}(t_{\rm end})|/16\pi^2$ $\ll 0.5$.

\section{Numerical results}

The running of ${\mbox{\boldmath $A$}}(t)$ depends strongly on the
behaviour of $\lambda(t)$. For small Higgs masses, the usual RG flow in
the SM leads to an instability of the EW vacuum caused by negative
values of $\lambda(t)$ in a certain range of $t$
\cite{Sher,espinosa}. The same happens in the presence of a
non-minimal curvature coupling.
\begin{figure}[h]
\includegraphics{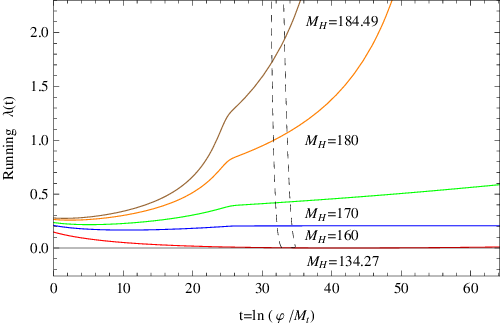}
\caption{Running $\lambda(t)$ for five values of the Higgs
mass above the instability threshold. Dashed curves mark the
boundaries of the inflation domain $t_{\rm end}\leq t\leq t_{\rm
in}$ \cite{RGHiggs}.}
 \label{Fig.1}
\end{figure}
The numerical solution for $\lambda(t)$ is shown in Fig.\ref{Fig.1}
for five values of the Higgs mass and the value of top quark mass
$M_t=171$ GeV. The lowest one corresponds to the boundary of the
instability window,
    \begin{equation}
    M_{\rm H}^{\rm inst}\simeq 134.27\; {\rm GeV},      \label{criticalmass}
    \end{equation}
for which $\lambda(t)$ bounces back to positive values after
vanishing at $t_{\rm inst}\sim 41.6$ or $\varphi_{\rm inst}\sim 80
M_{\rm P}$. The shape of the corresponding effective potential in the
Einstein frame is depicted in Fig.~\ref{Fig.2} and shows the existence
of a false vacuum at this instability scale.
\begin{figure}[h]
\includegraphics{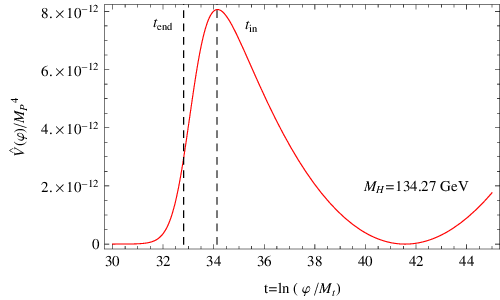}
\caption{\small The Einstein frame effective potential for the
instability threshold $M_{\rm H}^{\rm inst}=134.27$ GeV. A false
vacuum occurs at the instability scale $t_{\rm inst}\simeq 41.6$,
$\varphi_{\rm inst}\sim 80 M_{\rm P}$, which is much higher than the
Planck scale. A possible domain of inflation (ruled out by the lower
$n_s$ CMB bound) is again marked by dashed lines \cite{RGHiggs}. \label{Fig.2}}
\end{figure}
It turns out that the relevant $\xi(t)$ is nearly constant and is
about $5000$ (see below), so that the factor
(\ref{s}) at $t_{\rm inst}$ is very small, $s\simeq 1/6\xi\sim
0.00005$. Thus the situation is different from the usual SM
with $s=1$, and numerically the critical value turns out to be
higher than the known SM stability bound $\sim 125$ GeV
\cite{espinosa}.

Figure~1 shows that near the instability threshold $M_{\rm H}=M_{\rm
  H}^{\rm inst}$ the running coupling $\lambda(t)$ stays very small
for all scales $t$ relevant to the observable CMB. This follows from
the fact that the positive running of $\lambda(t)$
caused by the term $(18 s^2+6)\lambda^2$ in $\beta_\lambda$,
(see (\ref{beta-lambda})), is much slower for $s\ll 1$ than that of
the usual SM driven by the term $24\lambda^2$.

\begin{figure}[h]
\includegraphics{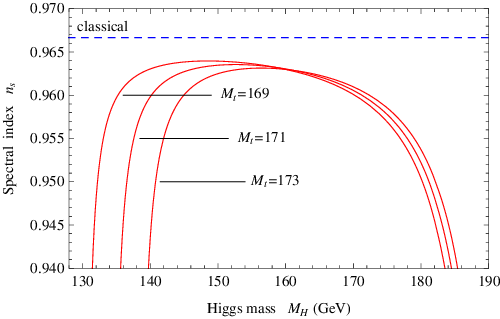}
\caption{ The spectral index $n_s$ as a function of the Higgs
mass $M_{\rm H}$ for three values of the top quark mass \cite{RGHiggs}.
 \label{Fig.4}}
\end{figure}

The RG running of ${\mbox{\boldmath $A_I$}}(t)$ explains the main
difference from the results of the one-loop calculations in
\cite{we}. ${\mbox{\boldmath $A_I$}}(t)$ runs from big negative values
${\mbox{\boldmath $A_I$}}(0)<-20$ at the EW scale to small
but also negative values at the inflation
scale below $t_{\rm inst}$. This makes the CMB data compatible with
the generally accepted Higgs mass range. Indeed, the knowledge of the
RG flow immediately allows one to obtain
${\mbox{\boldmath$A_I$}}(t_{\rm end})$ and $x_{\rm end}$ and thus to
find the parameters of the CMB power spectrum (\ref{ns})--(\ref{r}) as
functions of $M_{\rm H}$. The parameter of primary interest --
the spectral index -- is given by (\ref{ns}) with $x=x_{\rm end}\equiv
N{\mbox{\boldmath
$A_I$}}(t_{\rm end})/48\pi^2$ and depicted in Fig.~\ref {Fig.4}. Even
for low values of the Higgs mass above the stability bound, $n_s$
falls into the range admissible by the CMB constraint existing now at
the $2\sigma$ confidence level (based on the combined
WMAP+SPT+BAO+$H_0$ data \cite{WMAP,SPT}) $0.948 <n_s(k_0)<0.986$.

The spectral index drops below 0.95 only for large $x_{\rm end}<0$
or large negative ${\mbox{\boldmath $A_I$}}(t_{\rm end})$, which
happens only when $M_{\rm H}$ either approaches the instability bound or
exceeds 180 GeV at the decreasing branch of the $n_s$ graph. Thus
we get lower and upper bounds on the Higgs mass, which both
follow from the lower bound of the CMB data. Numerical analysis for
the corresponding $x_{\rm end}\simeq -1.4$ gives for $M_t=171$ GeV
the following range for the  CMB compatible Higgs mass:
    \begin{equation}
    135.6\; {\rm GeV}\leq M_{\rm H}
    \leq 184.5\; {\rm GeV}.       \label{CMBmass}
    \end{equation}
Both bounds belong to the nonlinear domain of
(\ref{ns}) with $x_{\rm end}\simeq-1.4$, but the quantity
$|\mbox{\boldmath $A_I$}(t_{\rm end})|/16\pi^2=0.07\ll 0.5$
satisfies the restriction mentioned above, and their calculation is
still in the domain of our linear in logs approximation.

The upper bound on $n_s$ does not generate restrictions on $M_{\rm H}$.
The lower CMB bound in (\ref{CMBmass}) is slightly higher than the
instability bound $M_{\rm H}^{\rm inst}=134.27$ GeV. In turn, this bound
depends on the initial data for weak and strong couplings and on the
top quark mass $M_t$, which is known with less precision. The above
bounds were obtained for $M_t=171$ GeV. Results for the neighboring
values $M_t=171\pm2$ GeV are presented in Fig.~\ref{Fig.4} to show
the pattern of their dependence on $M_t$.

\section{Gradient and curvature expansion cutoff and naturalness}

The expression (\ref{effaction}) is a truncation of the curvature
and derivative expansion of the full effective action. It was
repeatedly claimed that with large $\xi$ the weak field version of
this expansion on a flat (and empty) space background has a cutoff
$4\pi M_{\rm P}/\xi$ \cite{BurgLeeTrott,Barbon}. This scale is
essentially lower than the Higgs field during inflation $\varphi\sim
M_{\rm P}/\sqrt\xi$ and, therefore, seems to invalidate predictions based
on (\ref{effaction}) unless an unnatural suppression of
higher-dimensional operators is assumed. The attempt to improve the situation
by transition to the Einstein frame \cite{LernerMcDonald} was
claimed to fail \cite{BurgLeeTrott1,Hertzberg,Kaiser} for a
multiplet  Higgs field involving Nambu-Goldstone modes.

In the following, we show that these objections against naturalness are not
conclusive. First, as mentioned above, a large and almost constant value of $\varphi$
during inflation is not really indicative of a large energy scale
of the problem. In contrast to curvature and energy density, the
inflaton itself is not a physical observable, but rather a
configuration space coordinate of the model. Second, we now show
that the inflation scale actually lies below the gradient expansion
cutoff, and this justifies the naturalness of the obtained results. No
transition to another conformal frame is needed for this purpose, but rather
a resummation accounting for a transition to a large $\varphi$
background.

Indeed, the main peculiarity of the model (\ref{inf-grav}) is that
in the background field method with small derivatives the role of
the effective Planck mass is played by $\sqrt{M_{\rm P}^2+\xi\varphi^2}$.
Note that this effect is not a quantum one, it arises already at the tree level. 
The power-counting method of \cite{BurgLeeTrott} underlying the
derivation of the cutoff $4\pi M_{\rm P}/\xi$ also applies here, but with
the Planck mass $M_{\rm P}$ replaced by the effective one,
$M_{\rm P}\to\sqrt{M_{\rm P}^2+\xi\varphi^2}>\sqrt\xi\varphi$. The resulting
cutoff is thus bounded from below by
    \begin{equation}
    \Lambda(\varphi)=\frac{4\pi\varphi}{\sqrt\xi},  \label{cutoff}
    \end{equation}
and this bound can be used as a {\em running} cutoff of the gradient
and curvature expansion. The origin of this cutoff can be
demonstrated in the one-loop approximation. When calculated in the
{\em Jordan} frame, for the one-loop divergences quadratic in the
curvature $R$ the dominating  $\xi$ contribution is (this can be
easily deduced from Appendix of \cite{RGHiggs})
    \begin{equation}
    \xi^2\frac{R^2}{16\pi^2}.     \label{counterterm}
    \end{equation}
As compared to the tree-level part linear in the curvature $\sim
(M_{\rm P}^2+\xi\varphi^2)R$, the one loop $R^2$-term turns out to be
suppressed by the above cutoff factor
$16\pi^2(M_{\rm P}^2+\xi\varphi^2)/\xi^2\simeq\Lambda^2$.

The on-shell curvature estimate at the inflation stage reads
$R\sim V/U\sim\lambda\varphi^2/\xi$ in the Jordan frame,
so that the resulting curvature
expansion runs in powers of
    \begin{equation}
    \frac{R}{\Lambda^2}
    \sim\frac\lambda{16\pi^2}      \label{curvatureexpansion}
    \end{equation}
and remains valid in the usual perturbation theory range of SM, for which
$\lambda/16\pi^2\ll 1$. This works perfectly well in our Higgs
inflation model, because in the full CMB-compatible range of the
Higgs mass one has $\lambda<2$ (see Fig.~\ref{Fig.1}). The characteristic energy 
of quantum-gravitational fluctuations during inflation (sometimes referred as the Gibbons-Hawking temperature) $E \sim \sqrt{R} \sim \sqrt{\lambda}\varphi/\sqrt{\xi}$ is also 
much lower than the cutoff (34).  

From the viewpoint of the gradient expansion for $\varphi$, this cutoff
is even more efficient. Indeed, the inflaton field gradient can be
expressed in terms of the inflaton potential $\hat V$ and the
inflation smallness parameter $\hat\varepsilon$ taken in the Einstein
frame,
$\dot\varphi\simeq(\varphi^2/M_{\rm P}^2)
(\xi\hat\varepsilon\hat{V}/18)^{1/2}$. With $\hat V\simeq\lambda
M_{\rm P}^4/4\xi^2$, this immediately yields the
gradient expansion in powers of
    \begin{equation}
    \frac{\partial}{\Lambda}
    \sim\frac1\Lambda\frac{\dot\varphi}\varphi\simeq
    \frac{\sqrt\lambda}{48\pi}
    \sqrt{2\hat\varepsilon},                \label{partialexpansion}
    \end{equation}
which is even better than (\ref{curvatureexpansion}) by a factor
ranging from $1/N$ at the beginning of inflation to $O(1)$ at the end
of it.

Equations (\ref{curvatureexpansion}) and (\ref{partialexpansion}) justify
the effective action truncation in (\ref{effaction}) in the
inflationary domain. Thus, only multi-loop corrections to the
coefficient functions $V(\varphi)$, $U(\varphi)$, and $G(\varphi)$
 in the form of higher-dimensional
operators $(\varphi/\Lambda)^n$ may stay beyond control 
and violate the flatness of the
effective potential necessary for inflation. However, in view of the
form of the running cutoff (\ref{cutoff}) they might be large, but
do not affect the shape of these coefficient functions because of
the field independence of the ratio $\varphi/\Lambda$. Only the
logarithmic running of couplings in
(\ref{RGeffpot})--(\ref{phirenorm1}) controlled by the RG dominates the
quantum input in the inflationary dynamics and its CMB
spectra.\footnote{This is like the logarithmic term in (\ref{hatVbigphi}),
which dominates over the nearly flat classical part of the inflaton
potential and qualitatively modifies the tree-level predictions of the
theory \cite{we}.}

The attempt to justify this treatment was undertaken in \cite{BMSS} where the class of theories was formulated, in which the low-energy RG flow seems being protected from the arbitrariness of a UV completion. This is a class of models
possessing asymptotic scale invariance (or inflaton shift invariance in the Einstein frame) treated within the minimal subtraction scheme totally discarding power divergences. The authors of \cite{BMSS} claim that due to the asymptotic shift symmetry of the model in the Einstein frame the field-dependent cutoff at the inflation scale is much higher than (\ref{cutoff}) and is given by $\Lambda_E\sim \sqrt\xi\varphi$. This strongly supports the naturalness of Higgs inflation along with its consistency at the reheating and Big Bang stages and allows the authors of \cite{BMSS} to implement perturbation expansion in inflation smallness parameters, similar to (\ref{partialexpansion}). Unfortunately, a large value of their cutoff compared to (\ref{cutoff}) follows from the fact that in \cite{BMSS} it was identified from the violation of only tree-level unitarity, whereas the quantum corrections were estimated only in the scalar sector and specifically in the Einstein frame. 
The situation with gravitational counterterms and radiative corrections is trickier. It is only in this frame that the strongest curvature squared counterterm (when recalculated back to the Jordan frame metric) is $O(1)R^2/16\pi^2$ rather than (\ref{counterterm}). This nontrivial fact of frame dependence was discovered in the old paper \cite{our-ren}. Interestingly, a small $R^2$-counterterm appears also in the Jordan frame in the case of a scalar singlet, because the non-minimally coupled $N$-plet scalar field generates a one-loop counterterm $\sim\xi^2(N-1)R^2$ \cite{future}, but this miraculous cancelation for $N=1$ does not 
work for the physical four-component Higgs field.

\section{Conclusions and discussion}

The lower limit of the Higgs mass range compatible with the CMB data
for the spectral
index of the power spectrum, which we have obtained above in
(\ref{CMBmass}), is about 10 GeV higher than the
value of the Higgs mass announced at CERN
\cite{CERN}. As advocated in \cite{BezShap3}, the needed
decrease of
the lower bound on $M_{\rm H}$ can be achieved by the inclusion of
two-loop corrections. But as discussed in the Introduction, these
corrections are not really legitimate unless we resolve a number of
theoretical issues associated with the gauge-independent and
parametrization-independent definition of the cosmological
characteristics of the observable CMB. Thus, the result
(\ref{CMBmass}) should not be regarded as falsifying Higgs
inflation. Moreover, preliminary calculations show that two-loop
running of coupling constants indeed has a tendency to decreasing the
lower bound of (\ref{CMBmass}) like in \cite{BezShap3}. All this opens
interesting prospects for further studies of this model.

In this respect, a lot of work has to be done to reconsider the results
of \cite{Wil,BezShap3} usually referred to as two-loop
applications in Higgs inflation model. Despite a good agreement of
their conclusions with the currently pending range around $125$ GeV for
the Higgs mass, the variety of discrepancies between these works (and
also at the one-loop level with our paper \cite{RGHiggs}) brings up a
lot of concern about the source of these discrepancies and
ambiguities. Apart from pointing out to a general issue of gauge and
parametrization dependence discussed in the Introduction, we do not list
these ambiguities here.\footnote{One of the possible inconsistencies of
  the two-loop treatment in \cite{BezShap3} may be the fact that the
  two-loop effective potential is taken from the Cartesian
  coordinates calculation of \cite{Fordetal}, whereas the chiral phase
  of the SM employed in this work requires a spherical parametrization in
  terms of radial Higgs and angular Goldstone modes. Simple
  nullification of Nambu-Goldstone masses might be not enough for the
  transition from the result of \cite{Fordetal} to this
  parametrization.} Regarding the critique to be drawn on the
asymptotic safety approach to the EW stability of the Higgs inflation
model in \cite{BezShap4}, we might focus on the picture of the
effective potential in Fig.~\ref{Fig.2} at the critical value of the
Higgs mass. In contrast to anticipations that the gravitational
embedding of the SM affects the relation between the Fermi
scale and the Planck fixed point only weakly (a pivotal observation
that the scale
of this point with great precision turns out to be the Planck one
\cite{BezShap4}), we see from this figure that with a large nonminimal
$\xi\sim 5000$ the scale of this point is about $80 M_{\rm P}\gg 1
M_{\rm P}$. This means that this property, which served as a strong
motivation for this approach, is in fact very sensitive to the type of
the gravitational embedding, or at least strongly frame and
parametrization dependent. This gives additional motivation to
seriously reconsider this approach within explicit perturbation theory
calculations.

In conclusion, let us summarize the findings of our work. The result
(\ref{CMBmass}) is based on the RG improvement of the analytical
formalism in \cite{we}. A peculiarity of this formalism is that for
large $\xi\gg 1$ the effect of SM phenomenology on inflation is
universally encoded in one quantity -- the inflationary anomalous scaling
${\mbox{\boldmath $A_I$}}$. It was earlier introduced in
\cite{we-scale} for a generic gauge theory and is dominated in the SM
by contributions of heavy particles -- ($W^\pm$, $Z$)-bosons, top
quark, and Goldstone modes. The RG running raises a large negative EW
value of ${\mbox{\boldmath $A_I$}}$ to a {\em small negative value} at
the inflation scale. This ultimately leads to the range
(\ref{CMBmass}) for the  Higgs masses compatible with the CMB data.

This mechanism can be interpreted as asymptotic freedom, because
${\mbox{\boldmath $A_I$}}/64\pi^2$ determines the strength of quantum
corrections in inflationary dynamics \cite{BK,we}. Usually, asymptotic
freedom is associated with the asymptotic decrease of some running
coupling constants to zero. Here, this phenomenon is trickier because
it occurs in the interior of the range  and fails near its lower and
upper boundaries. Quantum
effects are small only in the middle part of (\ref{CMBmass}) with a
moderately small $\lambda$, where $n_s$ is close to the
``classical'' limit $1-2/N\simeq 0.967$ for $x\equiv
N{\mbox{\boldmath $A$}}/48\pi^2\ll 1$. Thus, the original claim of
\cite{BezShap1} on the smallness of quantum corrections is right, but this
smallness, wherever it takes place, is achieved via a RG summation of
big leading logarithms.

We have also demonstrated the naturalness of the gradient and
curvature expansion in this model, which is guaranteed within the
conventional perturbation theory range of SM, $\lambda/16\pi^2\ll 1$,
and which holds in the whole range of the CMB compatible Higgs mass
(\ref{CMBmass}) -- the latter property being a consequence of the
asymptotic freedom of
the above type. This result is achieved by the background field
resummation of weak field perturbation theory, leading to the
replacement of the fundamental Planck mass in the known cutoff $4\pi
M_{\rm P}/\xi$ \cite{BurgLeeTrott,Barbon} by the effective one. Partly
(modulo corrections to inflaton potential, which are unlikely to spoil
its shape), this refutes the objections of \cite{BurgLeeTrott,Barbon}
based on the analysis of scattering amplitudes in EW vacuum
background. The smallness of the cutoff in this background does not
contradict the physical bounds on the Higgs mass originating from the CMB data,
for the following reasons. The determination of $M_{\rm H}$ takes place,
of course, at the TeV scale, which is much below the non-minimal Higgs
cutoff $4\pi M_{\rm P}/\xi$, whereas inflationary dynamics and CMB
formation occur for $\lambda/16\pi^2\ll 1$ below the {\em running}
cutoff $\Lambda(\varphi)=4\pi\varphi/\sqrt\xi$. It is the phenomenon
of inflation which due to an exponentially large stretching connects these
two scales and allows one to probe the physics of the underlying
SM by CMB observations at the 500 Mpc wavelength scale.

Before summing up, let us formulate once again the basic assumptions
made in the present paper.
We have established a relation between observable cosmological data
(the spectral index $n_S$)
and data coming from particle physics. This relation arises due to
the fact that in the modified (scalar-tensor) gravity in the early Universe the classical Friedmann
evolution is essentially modified by quantum corrections. These
quantum corrections depend on interaction couplings of SM particles
with the Higgs field, which plays the role of the inflaton in the
model under consideration. To relate values of these couplings
measured at the electroweak scale with their hypothetical values at
the inflationary scale, we have used the renormalization group
formalism. We did this in spite of the well-known non-renormalizability
of quantum gravity  for two reasons. First, below a certain scale
one can use an effective field theory and all the participants of
the related discussions agree that this scale is not lower than
$M_{\rm P}/\xi$. Second, for large values of the scalar field (or,
in other words, at high energies), the theory possesses a scale
invariance (see also discussion of this point in \cite{BMSS,KS12}). 
It is this invariance which defends us from an
uncontrollable growth of quantum corrections. The question then
arises: is not  the transition between these two `safe' regions of values
of the scalar field dangerous? The hypothesis which we make consists
in the hope that the use of the continuous $s$-factor (see the
section 3 and the comments there) provides some kind of bridge
between these two regions, smoothly interpolating between low values
of the Higgs field, where the effective theory is valid, and high
values where the almost exact scale invariance is present.

To summarize, the inflationary scenario driven by the SM Higgs boson
with a strong non-minimal coupling to curvature looks very
promising. This model supports the hypothesis that an appropriately
extended SM can be a consistent quantum field theory all the way up to the
quantum-gravity regime and can in principle yield a fundamental
explanation of all major
phenomena in early and late cosmology \cite{nuMSM,dark}.
Further study of these issues can shed new light on the problem of the
range of validity of this model. Ultimately, it will be the
strongly anticipated confirmation of the recently announced discovery
of the Higgs particle at the LHC and the more precise determination of the
primordial spectral index $n_s$ by the Planck satellite that might
decide the fate of this model.

\begin{acknowledgements}
The authors are grateful to F. Bezrukov, M. Shaposhnikov, and O.
Teryaev for fruitful and thought-provoking correspondence and
discussions and also benefitted from discussions with D.~Diakonov,
I.~Ginzburg, N.~Kaloper, I.~M.~Khalatnikov, D.~V.~Shirkov,
S.~Solodukhin, G.~P.~Vacca, G.~Venturi, and R.~Woodard. A.B. is
especially thankful to E.~Alvarez, J.~Barbon, J.~Garriga, and
V.~Mukhanov for the discussion of naturalness in this model. A.B.
and A.K. acknowledge support by grant 436 RUS 17/3/07 of the
German Science Foundation (DFG) for their visit to the University of
Cologne. The work of A.B. was also supported by the RFBR grant
11-02-00512. A.K. and A.S. were
partially supported by the RFBR grant 11-02-00643.
The work of C.F.S.
was supported by the Villigst Foundation. A.B. acknowledges the
hospitality of LMPT at the University of Tours. A.S. also
acknowledges RESCEU hospitality as a visiting professor.
\end{acknowledgements}




\end{document}